# Analyzing P300 Distractors for Target Reconstruction

Jonathan R. McDaniel, *Member IEEE*, Stephen M. Gordon, *Member IEEE*, Amelia J. Solon, and Vernon J. Lawhern, *Member IEEE*

*Abstract*—P300-based brain-computer interfaces (BCIs) are often trained per-user and per-application space. Training such models requires ground truth knowledge of target and non-target stimulus categories during model training, which imparts bias into the model. Additionally, not all non-targets are created equal; some may contain visual features that resemble targets or may otherwise be visually salient. Current research has indicated that non-target distractors may elicit attenuated P300 responses based on the perceptual similarity of these distractors to the target category. To minimize this bias, and enable a more nuanced analysis, we use a generalized BCI approach that is fit to neither user nor task. We do not seek to improve the overall accuracy of the BCI with our generalized approach; we instead demonstrate the utility of our approach for identifying target-related image features. When combined with other intelligent agents, such as computer vision systems, the performance of the generalized model equals that of the user-specific models, without any user specific data.

## I. Introduction

One function of the human visual system is to identify task-relevant objects and place representations of those objects in working memory [1-2]. Electrophysiologically, it is believed that the P300 response performs a critical function along this pathway, specifically, the gating of low-level perception to higher-order memory [3]. Because the P300 response sits between initial perception and cognitive function, it is the focus of many studies in the context of perception and decision-making. The high signal-to-noise ratio of P300s also make it a popular input signal for electroencephalogram- (EEG) based brain-computer interfaces (BCIs).

P300-based BCIs are designed so that an image of interest, the "target", occurs relatively infrequently amongst an array of other "distractor" and "background" stimuli. The BCI's goal is to distinguish, via the presence of a P300, target stimuli from both distractor and background stimuli.

Both user and BCI system should thus ignore background and distractor stimuli. Background stimuli are perceptually distinct from target stimuli, whereas distractors are not. Although distractors differ from the target in a measurable way, they can elicit attenuated P300 responses [4-7]. To optimize performance (i.e. target detection), BCI designers typically fit models to both the user and the task (i.e. a particular target stimulus category).

While such an approach can be useful in laboratory environments, or for clinical populations, where stimuli are either 1) pre-labeled as target, distractor, or background or 2) drawn from a relatively small set of competing options, there remains many open questions about the utility of P300-based BCIs for healthy populations operating in complex environments. We investigate a relatively new form of P300 BCI, originally termed cortically-coupled computer vision (CCCV) [8]. In CCCV, the evoked potentials are paired with computer vision (CV) agents to improve the overall performance of the system. In this work, our BCI component is unique in that we intentionally do not fit our P300 BCI to the user or task to which it is applied. We show that such a generalized BCI is better at detecting the subtle response differences that can ultimately aid the CV agents that are responsible for making inferences about the target and target-related features.

A diagram of our sample system is shown in Fig. 1. The focus of this paper is the P300 BCI component (Fig. 1, dark grey). For our investigation, we analyze data from a 4 Hz rapid serial visual presentation (RSVP) task. In our experiment, subjects were instructed to look for a target stimulus amongst an array of distractors. The distractors varied in similarity to the target along several dimensions. For our P300 BCI, we use the deep learning approach described in [9]. We built our BCI for across-experiment and across-subject applications in order to remove the biases that result from explicitly treating all distractors as equal. When compared to models that were fit to the user and task, our approach shows 1) improved target detection, as well as 2) more accurate target features prediction using only distractor stimuli. These results demonstrate the value of generalized models and prove that these models are capable of revealing more information about the task than user/task-specific approaches.

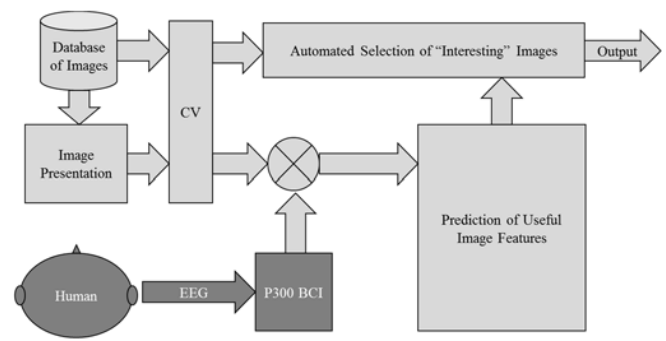

Figure 1. The P300 BCI uses EEG data to tag images as "interesting". A CV agent processes those images in order to identify commonalities along the different feature dimensions, so that an automated system can apply similar selection methods to a larger database of images.

*Research supported by the U. S. Army Research Laboratory.

J. R. McDaniel (phone: 571-227-6124; e-mail: jmcdaniel@dcscorp.com), S. M. Gordon (email: sgordon@dcscorp.com), and A. J. Solon (email: asolon@dcscorp.com) are with DCS Corp., Alexandria VA, 22310 USA.

V. J. Lawhern is with the U. S. Army Research Lab, Aberdeen MD, 21005 USA (e-mail: vernon.j.lawhern.civ@mail.mil).

## II. BACKGROUND

### A. Passive BCI

Starting with the first published BCI work [10], the goal of the majority of BCI research and development has been to restore some functionality in humans lost due to injury or disease. These functionalities can be broadly categorized as either restoring some ability to communicate, or providing motor activation and control. However, more recently, there has been growing interest in the development of so-called "passive" BCIs for healthy populations [11].

In contrast to conventional BCIs, the goal in passive BCIs is often to enrich a human-machine interaction, and does not require intentional input from the user. Prior passive BCI research has been used to assess cognitive and affective states like action preparation [12], error processing [13], and workload [14], and have used error-related potentials to correct the outcome of manually-controlled action [11].

### B. Cortically Coupled Computer Vision

The concept of CCCV was originally proposed by [8], where image triage is performed by a hybrid human/CV system. The CV system rapidly triages large volumes of images in order to identify those possessing specific features. The human user visually inspects images, and a BCI uses their neural responses to tag images deemed "interesting". To process a large volume of images, users are often shown images in rapid succession (i.e. RSVP). Additionally, a feedback loop provides CV-labeled data back to the user for evaluation. There are also other machine learning-based agents that manage and maintain linked representations of "interesting" image features. The net result is a hybrid system that outperforms both the CV and human at tagging relevant images of interest.

While CCCV is a tremendous leap forward in the application of P300-based BCIs for healthy populations, the system described in [8] is based on a BCI that is fit to the user and task. We argue here that by decoupling the BCI from the user/task, the system is better able to capture feature properties of the target category.

### C. Continuum of P300 Responses

Prior research has reported that non-target distractors—stimuli individuals were instructed to ignore—can produce intermediate P300s [4-7]. These non-target distractors are often perceptually or categorically related to the target. Thus, optimal task performance requires attentional processing of both targets and non-targets with the degree of similarity between the two influencing the level of attention allocation.

In [4], researchers found that distractors produced attenuated P300 responses and that the amplitude of the P300 was a function of target/distractor similarity. These results suggest that the "target-ness" of particular stimuli exists along a continuum of perceptual, or categorical, features.

## III. METHODS

### A. BCI Model Development

We built a generalized P300 detector using the EEGNet Deep Learning (DL) architecture [9]. Previously, we showed that EEGNet enabled cross-subject transfer performance equal to or better than conventional approaches for several BCI paradigms. EEGNet is also the model we used to obtain our cross-experiment results described in [15]. The architecture is composed of 16 spatial filters, implemented as a convolution across the channel dimension (Layer 1). Layers 2 and 3 of the network use spatio-temporal convolutions to learn correlations in time and across spatial filters. All convolutional layers used batch normalization, dropout, and $L_2$ regularization to mitigate overfitting. We trained our BCI using previously collected experiments from our database (Table 1). All data, including the test dataset described below, were collected with the approval of the Institutional Review Board of the U. S. Army Research Laboratory.

We based our selection of training data on previous results, which indicated that a diverse, balanced training set provided the most accurate (when averaged across multiple hold-out experiments) generalized model [15]. Each experiment required visual stimulus detection as the primary task in either a RSVP (i.e. stimulus-locked) or guided-fixation (i.e. fixation-locked) paradigm. Each experiment required participants to detect target stimuli and either 1) count the number of occurrences, or 2) press a button when a target appeared. Some experiments contained targets (T) and background (B) only, while others additionally contained distractors (D). All EEG data were collected with the BioSemi ActiveTwo system, using either 64 or 256 channels.

### B. Test Data Set

We collected data from 10 individuals performing a 4 Hz RSVP task, where the stimuli were images of cards from a standard 52-card deck. At the beginning of the experiment, each participant was shown a single image in which each of the cards were laid side-by-side, and were instructed to freely pick any card as their target card. EEG data were recorded using a 64-channel BioSemi ActiveTwo system, digitized at 2048 Hz. The data were referenced to average mastoids, down-sampled to 128 Hz, and bandpass filtered 0.3-50 Hz.

EEG data epochs associated with the presentation of each card were classified, and the resulting scores used to compute the Area Under the Curve (AUC) metric. AUCs were also computed for scores derived by weighting the standard output scores according to a causal prediction of three target card features: color, suit, and value. These feature-weighted scores were created for each feature independently, resulting in three sets of modified scores. We also evaluate the BCI's target feature prediction accuracy using distractor card images only.

TABLE I. LIST OF TRAINING DATA

| Event Lock | # Subs | # Samples (T vs D/B) | Response | Experiment Description |
|---|---|---|---|---|
| Fixation | 16 | 1,764; 17,654 | Button | Guided fixations with varying workload [16] |
| Stimulus | 16 | 10,512; 99,504 | Count or Button | 2 Hz RSVP with static or moving targets |
| Stimulus | 20 | 5,265; 53,951 | Button | 1 Hz RSVP before / after phys. exertion [17] |
| Fixation | 16 | 3,394; 31,610 | Button | Free-viewing target detection in urban setting |
| Stimulus | 18 | 12,965; 291,854 | Button | Varying target difficulty [18] |
| **TOTAL** | **86** | **33,900; 494,573** | --- | --- |

We compare the performance of our generalized BCI model to two well-proven user- and task-specific BCI models—Hierarchical Discriminant Components Analysis (HDCA) [19], and xDAWN [20]—which we trained using block-wise 5-fold cross-validation. As the focus of this work is the EEG aspect of CCCV, we assumed a CV system capable of perfectly identifying the color, suit, and value of each card.

*C. Target Feature Prediction*

Predictions of the target card's features were made after the presentation of each image. For a given feature (e.g. suit), the target's predicted feature class (e.g. heart) for the nth image was determined as follows: the mean classifier scores of cards belonging to each feature class from images 1–n were computed, and the class with the highest mean determined the predicted class. For example, if the mean score of the hearts cards that occurred in the first n images is higher than the mean scores of the other suits, heart is the target card's predicted suit.

*D. Feature-Weighted Classification Scores*

To compute a feature-weighted score for a particular image and feature, first, the target card's predicted feature class is determined as described above. If the feature class of the current card—derived from the CV system—does not match the target's predicted feature class, the original score associated with that image will be down-weighted, whereas if the classes match, that score will remain unchanged. For example, if the current prediction for the target card's suit is heart, then the current image's score will be decreased if it is not a heart. In the case of a mismatch, the weight was chosen to be the p value resulting from statistically testing the differences in the score means used to make the target feature class prediction. Thus, as confidence in the prediction grows, the scores from cards not matching that prediction will receive increasingly less weight.

A t-test was used to compare the differences in the mean scores of the red and black cards, and an analysis of variance (ANOVA) was used to test among the four suits and among the card values. Values were grouped into four perceptually similar classes: 2-5, 6-10, Jack-King, and Ace.

## IV. RESULTS

Fig. 2 shows the baseline results for 1) xDAWN, 2) HDCA, 3) our generalized DL model, and 4) feature-weighted DL. For this data set, xDAWN performs better than either HDCA or our initial DL with respect to detecting target stimuli. However, this performance difference goes away once we incorporate feature weights into our DL predictions. The features weights are, of course, derived from the DL responses to all images, not just targets.

To better understand what is happening, Fig. 3 (top row) shows the prediction accuracy for determining the feature value for each feature category. To obtain these results we averaged the individual predictions for each feature value for non-target images only and selected the feature value with the highest average. As more trials are included, our generalized model performs better at predicting each of the individual feature values than either HDCA or xDAWN. The generalized approach is able to predict the suit of the participant's card in 60-70% of cases, the color in 90% of cases, and face-value-category in 80% of cases.

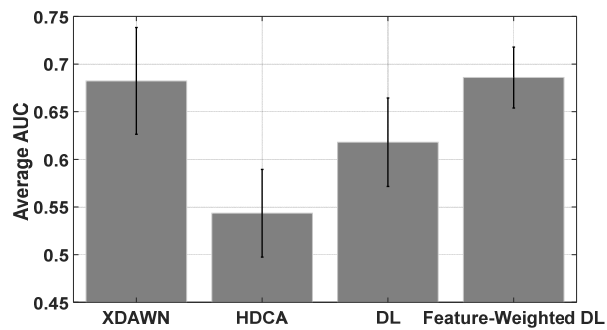

Figure 2. Average AUC for the user-specific and generalized approaches.

Fig. 3 (bottom row) shows AUC values for each approach as a function of the number of trials included. As can be seen from the figure, the performance of our DL model increases regardless of the feature used to weight the prediction. When using the color and value features, this generalized approach performs just as well at detecting target stimuli as the best user-specific approach (xDAWN).

The user-specific approaches do not see a uniform improvement in performance with including feature weights (some features improve performances while others do not). Here, we start our analysis at 200 trials to provide sufficient representation of target and non-target stimuli when computing AUC. Furthermore, we note that these results are biased towards the user-specific approaches since the predictions for these methods were computed in a cross-validation manner. The purpose of this figure is simply to present how the feature weights could improve performance for the user-specific models.

## V. DISCUSSION

By using a generalized BCI, we freed our system from some of the biases that result from treating all distractors as the same. As a result, we were able to take advantage of the continuous aspect of the P300 response to distractors and targets. When paired with labels derived from a hypothetical CV agent capable of perfect image feature extraction, we were able to correctly identify target feature values at a rate greater than task- and user-specific models. We were even able to identify these feature values without access to any instances of the target stimulus. We believe that the performance we have observed is because the BCI was forced to learn a generalized response by learning across multiple individuals and multiple distinct experimental paradigms, even though during training, the BCI was still given strictly labeled training data.

Obviously, our DL model was trained with significantly greater amounts of training data than our user-specific models, and we would generally expect that a model trained with more data would outperform a model that was trained with less data. That our generalized BCI did not originally outperform the user-specific models speaks to the power of incorporating user-specific data into the BCI. Thus, one avenue for future work is to investigate how the outputs between the generalized approach, which performed best at detecting the "target-ness" of non-target distractors, can be paired with a user-specific approach that performs better at detecting isolated targets. Such insight could have a tremendous positive impact on CV systems that must make inferences about image features and target stimuli. Rather than learning from just a few single target images, these systems could consider the entirety of images presented to the user by assessing the neural response along various dimensions of image features.

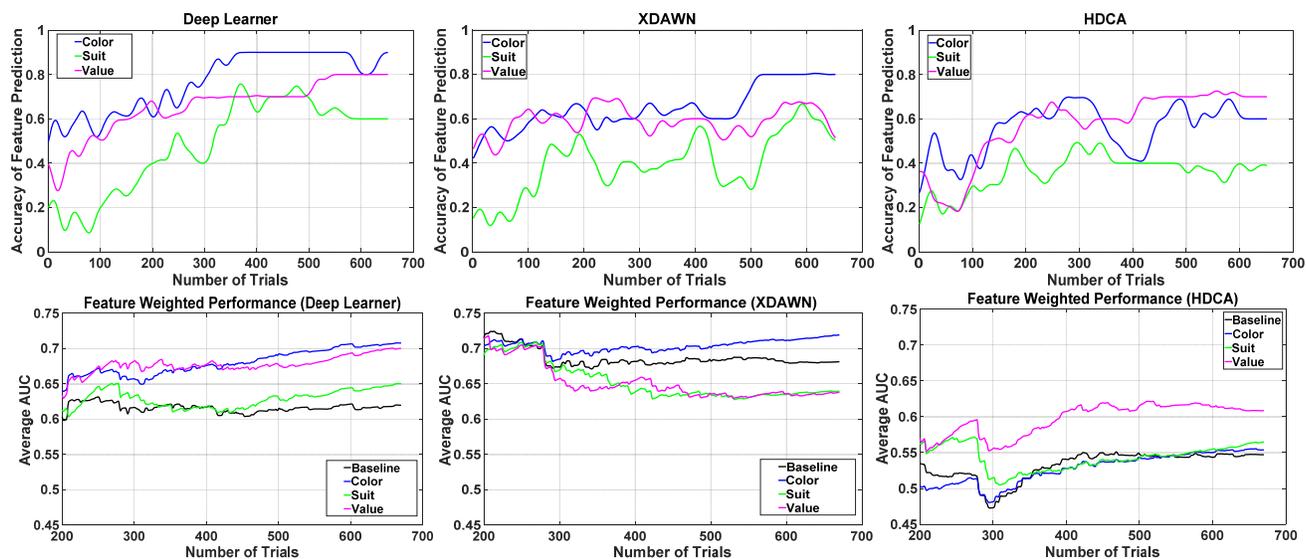

Figure 3. Top Row: Feature prediction accuracy—using only non-target trials—of the deep learner, xDAWN, and HDCA classifiers. Bottom Row: AUCs for classifier scores modified by the feature-based weights, computed as a function of the number of trials analyzed.

ACKNOWLEDGEMENT

This research was sponsored by the Army Research Laboratory under ARL-74A-HRCYB and through Cooperative Agreement Number W911NF-10-2-0022. The views and conclusions contained in this document are those of the authors and should not be interpreted as representing official policies, either expressed or implied, of the Army Research Laboratory or the U.S. Government. The U.S. Government is authorized to reproduce and distribute reprints for Government purposes notwithstanding any copyright notation herein.